\documentstyle[preprint,aps]{revtex}

\begin{document}
\title{Hamiltonian field theoretical model for a light quark condensate }
\author{A. Miranda}
\date{23rd August 2006}
\maketitle

\begin{abstract}
I propose an alternative Hamiltonian field theoretical model for a light
quark condensate that is compatible with QCD in the deep infrared. Key
electroweak data on flavourless pseudoscalar mesons are used for necessary
renormalizations. Light quark inertial masses are redefined in a new and
broader theoretical context.

PACS. : 11.30.Rd, 12.15.Ff, 14.65.Bt

{\it key words: }fermion condensates, light quarks, chiral symmetry, quantum
mechanical symmetry breaking, effective Hamiltonians.
\end{abstract}

\section{Introduction}

The aim of this paper is to reconsider again the old pre-QCD
Nambu-Jona-Lasinio(NJL) ideas, but this time in the light of the new
insights gained in the last 10 years. We shall make appropriate use of
contemporary methods and ideas on boson and fermion condensates that proved
to be so useful in both nuclear physics and condensed matter physics. The
phenomena associated with broken chiral-flavour symmetries and chiral
anomalies are indeed to be ultimately related to the physical nature and
structure of a presumed quark-gluon quantum vacuum.

We assume the existence of a u,d fermion condensate and work from there,
using mathematical methods more or less common to all fermionic systems at
zero temperature. Our basic degrees of freedom are chiral QCDu,d,s-quarks
but only in colour singlet combinations. Their couplings to the leptons and
gammas are assumed to be primarily as prescribed by the Standard Model (SM).

Updating the NJL idea, we design an effective Hamiltonian to be worked out
under the assumption that it has a stable Dirac-Hartree-Fock-Bogoliubov
(DHFB) state as its (approximate) ground state . \ Necessary
renormalizations are carried out using experimental meson masses and
electroweak data. Applications and detailed numerical results will be
published elsewhere.

\section{The relevant effective Hamiltonian}

We begin by making the standard symmetry assumptions[1,2,3]. The internal
symmetry that phenomenology suggests as the most relevant to the analysis of
the lightest quark sector is the chiral-flavour symmetry of the form

\[
G:SU_{NL}\otimes SU_{NR}\otimes U_{1L}\otimes U_{1R}\qquad (1) 
\]
with N=2,3. The chiral-flavour symmetry G undergoes various quantum
mechanical symmetry breakings \ [2,3] that reduces it down to flavour groups

\[
G\Longrightarrow H:SU_{NL+R}\otimes U_{1L+R}\qquad (2) 
\]

Thus G is a symmetry of the Hamiltonian (or equations of motion) whereas H
is a symmetry of the quantum vacuum. There are two well-understood kinds of
symmetry breakings that achieve this, both traceable to the quantum
vacuum[2,3,4]. The first kind is due to the existence of a specific
classically definable Landau-like vacuum long range order (LRO)[5], to be
defined more precisely in the following sections. It breaks the chiral $%
SU_{N}$ ($SU_{N}$) part of the algebra (1), leading to the emergence of the
right set of Goldstones(viz. \ 3 and 8). The other kind of quantum
mechanical symmetry breaking removes the axial $U_{1L-R}$ symmetry of the
Hamiltonian, as a consequence of the axial anomaly. The latter can be
interpreted as simply a consequence of the physical existence of the Dirac
vacuum[4]. This implies that there are no chiral doublets in the hadronic
phase of QCD in which we live. This information is the input used in
designing the effective Hamiltonian of this paper.

We choose our basic degrees of freedom to be included in this effective
Hamiltonian: these are (we use Bjorken and Drell notations, definitions and
conventions [6])

\[
(u_{L},d_{L})\oplus (u_{R},d_{R})\oplus (s_{L},s_{R})\qquad (3) 
\]
and their antiparticles.We assume input mass matrices:

\[
M_{0n}=\left( 
\begin{array}{ccc}
m_{u} & 0 & 0 \\ 
0 & m_{d} & 0 \\ 
0 & 0 & m_{s}
\end{array}
\right) \qquad (4) 
\]
Thus the simplest non-trivial relevant effective Hamiltonian can be assumed
to be

\[
{\bf H}_{st}=U_{0}+{\bf H}_{0}+{\bf V}^{(1)}+{\bf V}^{(2)}\qquad (5) 
\]

\[
{\bf H}_{0}=\sum_{n=cu,cd,cs}\int d^{3}\vec{x}\bar{q}_{nR}(\vec{x})(-i\vec{%
\gamma}.\vec{\nabla})q_{nR}(\vec{x}))+L\Longleftrightarrow R\qquad (6) 
\]

\[
{\bf V}^{(1)}=\sum_{n=cu,cd,cs}\int d^{3}\vec{x}\bar{q}_{nR}(\vec{x}%
)M_{0n}q_{nR}(\vec{x}))+L\Longleftrightarrow R\qquad (7) 
\]

\[
{\bf V}^{(2)}=\frac{g_{SP}}{8\pi \Lambda _{\chi }^{2}}\sum_{n,n^{\prime
}=cu,cd,cs}\int d^{3}\vec{x}\bar{q}_{nL}(\vec{x})q_{nR}(\vec{x})\bar{q}%
_{n^{\prime }R}(\vec{x})q_{n^{\prime }L}(\vec{x})+L\Longleftrightarrow R\
\qquad (8) 
\]

\[
+\frac{g_{VA}\ }{8\pi \Lambda _{\chi }^{2}}\sum_{n,n^{\prime }=cu,cd,cs}\int
d^{3}\vec{x}\bar{q}_{nL}(\vec{x})\gamma ^{\mu }q_{nL}(\vec{x})\bar{q}%
_{n^{\prime }L}(\vec{x})\gamma _{\mu }q_{n^{\prime }L}(\vec{x}%
)+L\Longleftrightarrow R\ \qquad (9) 
\]
The $q_{nL,R}$ are chiral field operators\ [6].The necessary inputs for this
model are:

(i) a real flavour diagonal $3\times 3$ ''input mass matrix'' M$_{0n}$;

(ii) Two real independent effective dimensionless couplings $g_{SP}$ and $%
g_{VA}$;

(iii) two fundamental mass scales, provided by an explicit $\Lambda _{\chi }$
$\sim 1GeV$ (related to the scale where massless quarks become massive
quarks) and an implicit $\Lambda _{QCD}\sim 300MeV$ (a parameter basically
fixing the overall size of physical hadrons). In the context of a non-chiral
quark model this region would give the medium range q\={q} potential.

\section{The relevant LRO}

Let us {\it define }the Landau-like long range order (LRO) parameter
appropriate to an ''u,d-quark condensate'' by {\it assuming} the existence
of a {\it robust} spinless, colourless and flavourless non-vanishing scalar
LRO[2,3]

\[
\Delta =\frac{g_{SP}}{8\pi \Lambda _{\chi }^{2}}\sum_{n=cu,cd,cs}<0|(\bar{q}%
_{nL}(\vec{x})q_{nR}(\vec{x}))|0>+L\Longleftrightarrow R\qquad (10) 
\]
Translation invariance ensures independence on space coordinates. This {\it %
defines} g$_{SP}$.The notation $|0>$ as used here should be explained. In a
many-body \ context the symbol 
\mbox{$\vert$}%
0%
\mbox{$>$}%
would usually mean that one is referring to a certain state vector in
Hilbert space representing the true ground state of the many-body system. In
the context of this paper, such a true ground state of QCD-quark-gluon
coupled fields is of course not only unknown but it is also irrelevant. This
notation is nevertheless adopted here, but $|0>$ merely defines a ''no
particle state'' which is just a tautology for ''normal operator
products''[7]. It has therefore nothing to do with any true physical ground
state. We shall sometimes refer to it rather loosely as the '' quark
vaccum''.

We work exclusively with approximate Heisenberg operators that play the role
of ''physical states''[7]. It is established that only colour singlets would
qualify as such. We try to find stationary solutions to the Heisenberg
equations of motion for these ''physical states''. Thus our ''no particle
state'' is nothing but a tautology for normal ordering DHFB quasiparticle
operators.

In order to include this assumption in the effective Hamiltonian we begin by
making a straightforward Boguliubov transformation using the Nambu-Gorkov
representation for the u,d QCD-quarks[5]:

\[
\left( 
\begin{array}{c}
\alpha _{n\lambda }(\vec{p}) \\ 
\beta _{\bar{n}\bar{\lambda}}^{+}(-\vec{p})
\end{array}
\right) =\sum_{h=\pm \frac{1}{2}}\left( 
\begin{array}{cc}
\sin \varphi _{nh\lambda }(p) & \cos \varphi _{nh\lambda }(p) \\ 
-\cos \varphi _{nh\lambda }(p) & \sin \varphi _{nh\lambda }(p)
\end{array}
\right) \left( 
\begin{array}{c}
b_{nh}(\vec{p}) \\ 
d_{\bar{n}\bar{h}}^{+}(-\vec{p})
\end{array}
\right) \quad \qquad (11) 
\]
where $\lambda =L,R$ are chiralities and h are helicities. The ''Bogoliubov
angles $\varphi _{nh\lambda }(p)$'' serve as adjustable variational
parameters. By separating out the bilinear from the non-bilinear terms we
find that

\[
{\bf H}_{st}(u,d,s)=U_{0}^{\prime }+{\bf H}_{0}^{\prime }+{\bf V}^{\prime
}\qquad (12) 
\]
where

\[
{\bf H}_{0}^{\prime }=\sum_{n=cu,cd,cs}\int d^{3}\vec{p}:\left( 
\begin{array}{cc}
\bar{q}_{nR}(\vec{p}) & \bar{q}_{nL}(\vec{p})
\end{array}
\right) \left( 
\begin{array}{cc}
|\vec{p}|\  & -\Delta _{n} \\ 
-\Delta _{n} & -|\vec{p}|
\end{array}
\right) \left( 
\begin{array}{c}
q_{nL}(\vec{p}) \\ 
q_{nR}(\vec{p})
\end{array}
\right) \qquad (13) 
\]
and V' represents the remaining (quadrilinear) terms.The Bogoliubov angles
are chosen so that ${\bf H}_{0}^{\prime }$ is fully diagonalized:

\[
{\bf H}_{0}^{\prime }=-U_{0}^{\prime }+\sum_{n=cu,cd,cs}\sum_{h=\pm \frac{1}{%
2}}\int d^{3}\vec{p}E_{n}(p)[b_{nh}^{+}(\vec{p})b_{nh}(\vec{p})+d_{\bar{n}%
\bar{h}}^{+}(-\vec{p})d_{\bar{n}\bar{h}}(-\vec{p})]\qquad (14) 
\]

\[
\sin ^{2}\varphi (p)=\frac{1}{2}(1-|\vec{p}|/E_{u,d}(p)))\qquad n=u,d\qquad
(15) 
\]

\[
E_{u,d}(p)=\sqrt{|\vec{p}|^{2}+(m_{u,d}+\Delta )^{2}}\qquad (16) 
\]

\[
E_{s}(p)=\sqrt{|\vec{p}|^{2}+m_{s}^{2}}\qquad (17) 
\]
where by definition

\[
b_{nh}(\vec{p})|0>=0=d_{\bar{n}\bar{h}}(-\vec{p})|0>\qquad (18) 
\]
Note the independence of the Bogoliubov angles on helicities /chiralities
that follow from the definition of the LRO. We shall refer to this as the
Dirac-Hartree-Fock-Bogoliubov(DHFB) static approximation.

\section{Renormalizations}

We shall have to renormalize the above theory, selecting experimental data
on pions and the etas for doing so[1]. These mesons, just like any physical
mesons, are here considered to be just complex poles of the physical
S-matrix amplitudes.

The electromagnetic sector of this model is represented by the Hamiltonian

\[
{\bf H}_{ew}={\bf H}_{0l\gamma }+{\bf V}_{ew}\qquad (19) 
\]
and will be treated perturbatively.${\bf H}_{0l\gamma }$ stands for free
leptons and gammas. The ${\bf V}_{ew}$ is given by the SM[2,3].

(i) Consider the $\pi ^{\pm }$main decay channel in the rest frame ,
defining the decay constant $f_{\pi ^{\pm }}$

\[
A(\ \pi ^{\pm }\longrightarrow \mu ^{\pm }+\nu _{\mu }(\bar{\nu}_{\mu
}))\equiv <0|J_{weak}|\pi ^{\pm }(M_{\pi ^{\pm }})>\qquad (20) 
\]
where

\[
|\pi ^{\pm }(M_{\pi ^{\pm }})>=|\pi ^{+}>\equiv \frac{1}{\sqrt{6}}%
\sum_{c}\sum_{h}\int d^{3}\vec{p}\Psi _{\pi ^{\pm }}(\vec{p})\ast
b_{cuh}^{+}(\vec{p})d_{\bar{c}\bar{d}\bar{h}}^{+}(-\vec{p})|0>\ (21) 
\]

\[
|<0|J_{weak}|\pi ^{\pm }(M_{\pi ^{\pm }})>|=|f_{\pi ^{\pm }}|\sqrt{\frac{%
M_{\pi ^{\pm }}}{2(2\pi )^{3}}}\ (22) 
\]
we find the condition

\[
\sqrt{\frac{M_{\pi ^{\pm }}}{3(2\pi )^{3}}}\frac{|f_{\pi ^{\pm }}|}{4\cos
\theta _{C}}=|\int \frac{d^{3}\vec{p}}{(2\pi )^{3}}\Psi _{\pi ^{\pm }}(\vec{p%
})\ast \cos 2\varphi (p)|\ (23) 
\]

(ii) Consider the $\pi ^{\pm }$charge radius, defined through

\[
<\pi ^{+}(\vec{p}_{2})|J_{em}^{\mu }(0)|\pi ^{+}(\vec{p}_{1})>=G_{\pi
}((p_{1}-p_{2})^{2})\frac{(p_{1}+p_{2})^{\mu }}{\sqrt{(2\pi )^{3}2E_{\pi
}(p_{2})(2\pi )^{3}2E_{\pi }(p_{1})}}\ (24) 
\]

where

\[
G_{\pi }((p_{1}-p_{2})^{2})=1+\frac{1}{6}<r_{\pi }^{2}>(p_{1}-p_{2})^{2}\
(25) 
\]

A simple estimate of the theoretical charge radius can be obtained by making
a reasonable ansatz for the (common) normalized internal wavefunction of $%
\pi ^{\pm }$:

\[
\Psi _{\pi ^{\pm }}(\vec{\rho};x,y)=\frac{1}{\sqrt{4\pi }}N(x,y)\exp (-\frac{%
1}{2}(\rho -x\Lambda _{\chi }^{-1})^{2}/(y\Lambda _{QCD}^{-1})^{2}\ (26) 
\]
where x,y are dimensionless variational parameters. So

\[
<r_{\pi }^{2}>=\Lambda _{\chi }^{2}\frac{F_{4}(x,y;\Lambda _{\chi }/\Lambda
_{QCD})}{F_{2}(x,y;\Lambda _{\chi }/\Lambda _{QCD})}\qquad (27) 
\]

\[
F_{j}(x,y;\Lambda _{\chi }/\Lambda _{QCD})=\int_{-x/y}^{\infty }du\exp
(-(\Lambda _{\chi }/\Lambda _{QCD})^{2}u^{2})(x+yu)^{j}\ (28) 
\]

(iii)Next, consider the main decay channel of the $\pi ^{0}$at rest:

\[
A(\pi ^{0}\longrightarrow \gamma (\vec{k})+\gamma (-\vec{k})=<\gamma (\vec{k}%
),\gamma (-\vec{k})|{\bf V}_{em}|\pi ^{0}>\qquad (29) 
\]

\[
|\pi ^{0}>=\frac{1}{\sqrt{12}}\sum_{c}\sum_{q=cu,cd}\sum_{h}\int d^{3}\vec{p}%
\Psi _{\pi ^{0}}(\vec{p})b_{cqh}^{+}(\vec{p})d_{\bar{c}\bar{q}\bar{h}}^{+}(-%
\vec{p})|0>\qquad (30) 
\]
But assuming the initial pseudoscalar at rest, we find that the absorptive
part of the (dominant) underlying Lorentz and gauge invariant q\={q}
annihilation amplitude $a$ (with massive quarks) is :

\[
k^{2}%
\mathop{\rm Im}%
a(\pi ^{0}(\Longleftrightarrow q+\bar{q})\longrightarrow \gamma (\vec{k}%
)+\gamma (-\vec{k}))\sim 
\]

\[
\sim \left( \Delta /k\right) ^{2}(1-\left( \Delta /k\right) ^{2})^{-\frac{1}{%
2}}\ln [1-(1-\left( \Delta /k\right) ^{2})^{\frac{1}{2}}/(1+(1-\left( \Delta
/k\right) ^{2})^{\frac{1}{2}})]\qquad (31) 
\]

.As $\Delta ,k\longrightarrow 0$ it can be shown that the lhs of (31) tends
to $\delta (k^{2})$ [8], contrary to na\"{i}ve expectations.The deep reason
for this is the existence of the fixed anomaly pole at $k=0$, a necessary
feature if the U$_{1em}$ gauge invariance is to be maintained. This pole
lives below the physical threshold( at about $k=2\Delta $), which can be
related to the physical mass of the $\pi ^{0}$ through the
Gell-Mann-Oakes-Renner formula [2,3]. The existence of this pole could be
guessed from (31) as the rise of \ \ $k^{2}%
\mathop{\rm Im}%
a$ as $k$ descends from infinity towards the physical threshold. This
provides another condition on our parameters.

(iv)In order to calculate the value of the $\eta _{0}-\eta _{0}^{\prime }$
anomaly [2,3]we shall have to include further interaction terms from (12).We
shall consider the effect only to leading order and ignore all isospin
mixings. Let us define

\[
|\eta _{i}>=\frac{1}{\sqrt{3}}\sum_{c}\sum_{h}\sum_{f_{i}}\int d^{3}\vec{p}%
\Psi _{\eta _{i}}(\vec{p})b_{cf_{i}h}^{+}(\vec{p})d_{\bar{c}\bar{f}_{i}\bar{h%
}}^{+}(-\vec{p}))|0>\ i=1,2\qquad (32) 
\]
with f$_{1}=u,d$ and f$_{2}=s$.

So by diagonalizing the $2\times 2$ matrix

\[
M_{ij}=<\eta _{i}|{\bf H}_{st}(u,d,s)-U_{0}|\eta _{j}>\qquad (33) 
\]
we find that the eigenvectors (in the approximation of keeping only the
q\={q} pair exchange diagrams) are

\[
|\eta _{0}>=\cos \theta |\eta _{1}>+\sin \theta |\eta _{2}>\qquad (34) 
\]

\[
|\eta _{0}^{\prime }>=\sin \theta |\eta _{1}>-\cos \theta |\eta _{2}>\qquad
(35) 
\]

where the mixing angle is given by

\[
\tan \theta =(M_{\eta ^{0}}-M_{11})/M_{12}=M_{21}/(M_{\eta _{0}^{\prime
}}-M_{22})\qquad (36) 
\]

\[
M_{11}=2E_{u,d}+|F_{1}|^{2}\qquad M_{22}=2E_{s}+|F_{2}|^{2}\qquad
M_{12}=F_{1}^{\ast }F_{2}+F_{1}F_{2}^{\ast }=M_{21}\qquad (37) 
\]

\[
F_{1}\equiv \frac{1}{\sqrt{12}}\int_{0}^{\infty }d^{3}\vec{p}\Psi _{\eta
_{1}}(\vec{p})A_{1}(\Delta ,\vec{p})\quad F_{2}\equiv \frac{1}{\sqrt{6}}%
\int_{0}^{\infty }d^{3}\vec{p}\Psi _{\eta _{2}}(\vec{p})A_{2}(M_{s},\vec{p}%
)\qquad (38) 
\]

\[
A_{1}(\Delta ,\vec{p})\equiv 12\Delta /\sqrt{|\vec{p}^{2}|+\Delta ^{2}}%
\qquad (39) 
\]

\[
A_{2}(M_{s},\vec{p})=6M_{s}/\sqrt{|\vec{p}^{2}|+M_{s}^{2}}\qquad (40) 
\]

The angle $\theta $ is thus part of our renormalized parameters.

\section{Conclusions}

We presented an Hamiltonian field theory in a DHFB-static approximation as a
complement/alternative to the conventional theory of pions and etas.
However, our theory (based on an updated version of the NJL field theory[9])
can be easily extended and has a much broader scope. It can also easily be
linked to non-chiral quark models and QCD in the deep infrared. All
renormalizations procedures are build in the theory itself and have of
course only meaning in the context of this theory, as every renormalization
scheme does in its own context. The issue of confinement, though not
essential to our case, is bypassed in a natural way. Detailed numerical fits
to experimenta data in order to get the renormalized parameters will be
published elsewhere.

REFERENCES

[1] \ \ Particle Data Group, http//pdg.lbl.gov/.

[2] \ S.Weinberg, in{\it \ The Quantum Theory of \ \ }

\ \ \ \ \ \ \ {\it Fields,Vol.1}(Vol.1(pp.62-81;213-229 ) and Vol.2(pp.
182-192)

\ \ \ \ \ \ \ edited by Cambridge University Press, U.K.1995-6.

[3] \ \ J. Donnoghue, E.Golowich and B.R.Holstein, in {\it Dynamics of the
Standard}

\ \ \ \ \ \ {\it \ Model }edited by Cambridge University Press,
U.K.1992,pp.157-183.

[4] \ R.Jackiw, hep-th/9903255 preprint, 1999.

[5] \ G.Volovik , in {\it The Universe in a helium Droplet}(pp.65-70){\it ,}
edited by

\ \ \ \ \ \ \ \ Clarendon Press, Oxford, 2003.

[6] \ \ J.Bjorken and S.Drell , in \ {\it Relativistic Quantum Fields, }%
edited by \ \ \ 

\ \ \ \ \ \ \ Mc-Graw-Hill Book Company,1965,pp.43-67.

[7] \ P. A.M.Dirac, in {\it Lectures in Quantum Field Theory, }edited by
Elfer Graduate

\ \ \ \ \ \ \ School, Yeshiva University, New York 1966.

[8] \ \ A.Dolgov and V.I. Zakharov, Nucl.Phys.B27,(1971)525.

[9] \ \ M. Buballa, hep-ph/0402234 preprint, 2004.

\end{document}